\def\eop{E^{\rm obs}_{\rm peak}}
\def\eiso{E_{\rm iso}}
\def\xiso{X_{\rm iso}}
\def\ep{E_{\rm peak}}
\def\einf{E^{\rm inf}_{\gamma}}
\def\egt{E^{\rm true}_{\gamma}}
\def\muv{\mu_{\rm v}}
\def\thetav{\theta_{\rm v}}
\def\thetan{\theta_{0}}
\def\thetaj{\theta_{\rm j}}

      [1999/12/01 v1.4c Il Nuovo Cimento]
\documentclass{cimento}

\usepackage{graphicx}
\title{Unifying XRFs and GRBs with a Fisher-Shaped Universal Jet Model}
\author{T. Q.~Donaghy\from{uofc},
C.~Graziani\from{uofc} \&
D. Q.~Lamb\from{uofc}}
\instlist{\inst{uofc} Dept. of Astronomy \& Astrophysics, 
University of Chicago\\ 5640 S. Ellis Ave., Chicago, IL, 60615}
\PACSes{\PACSit{98.70.Rz}{gamma-ray sources; gamma-ray bursts}}
\newcommand{\beq}{\begin{equation}}
\newcommand{\eeq}{\end{equation}}
\begin{document}

\maketitle

\begin{abstract}

We show analytically that GRB jets with an emissivity profile given by
the Fisher distribution, $\epsilon(\theta) = A \cdot \exp (B\cdot
\cos\theta)$, have the unique property of producing equal numbers of
bursts per logarithmic interval in $E_{\rm iso}$, and therefore in most
burst properties.  Since this broad distribution of burst properties is
a key feature found by HETE-2, a Fisher-shaped universal jet model can
explain many of the observed properties of XRFs, X-ray-rich GRBs, and
GRBs reasonably well, in contrast to a power-law universal model.  For
small viewing angles, the Fisher distribution can be approximated by a
Gaussian, whose properties have been explored by \cite{zhang2004}.  We
also show that the Fisher universal jet model produces a broad
distribution in the inferred radiated energy $E_{\gamma}^{\rm inf}$, in
contrast to the narrow distribution predicted by the uniform variable
opening-angle jet model \cite{lamb2005b}.  Here we present Monte Carlo
simulations of both a Fisher-shaped universal jet model and a
Fisher-shaped variable opening-angle jet model.

\end{abstract}

\vspace{-0.4truein}
\section{Introduction}

The HETE-2 results show that the properties of XRFs \cite{heise},
X-ray-rich GRBs, and GRBs form a continuum in the [$S_E(2-400~{\rm
kev}),\eop$] plane \cite{sakamoto2004b}.  They also show that the
relation between the isotropic-equivalent burst energy $\eiso$ and the
peak energy $\ep$ of the burst spectrum in $\nu F_\nu$ in the rest
frame of the burst found by \cite{amati2002} extends to XRFs and
X-ray-rich GRBs \cite{lamb2005}.  A key feature of the distribution of
bursts in these two planes is that the density of bursts is roughly
constant along these relations, implying equal numbers of bursts per
logarithmic interval in $S_E$, $\eop$, $\eiso$ and $\ep$.  These
results, when combined with earlier results \cite{heise,kippen2002},
strongly suggest that all three kinds of bursts are the same
phenomenon.  It is this possibility that motivates us to seek a unified
jet model of XRFs, X-ray-rich GRBs, and GRBs.

In our previous paper \cite{lamb2005b}, we explored two different
phenomenological jet models: a variable jet opening-angle model in
which the emissivity is uniform across the surface of the jet and a
universal jet model in which the emissivity is a power-law function of
the angle relative to the jet axis.  We showed that while the variable
jet opening-angle model can account for the observed properties of all
three kinds of bursts, the power-law universal jet model cannot easily
be extended to account for the observed properties of both XRFs and
GRBs. In response to that conclusion, \cite{dz2004,zhang2004}
considered a quasi-universal Gaussian jet model \cite{zhang2002}.  They
showed that such a model can explain many of the observed properties of
XRFs, X-ray-rich GRBs, and GRBs reasonably well.

Here we consider a universal jet model in which the emissivity of the
jet as a function of viewing angle is a Fisher  distribution (such a
distribution is the natural extension of the Gaussian distribution to a
sphere).  We show that the Fisher distribution has the unique property
that it produces equal numbers of bursts per logarithmic interval in
$\eiso$, and therefore in most burst properties, consistent with the
HETE-2 results.  We also show that the Fisher universal jet model
produces a broad distribution in the inferred radiated energy $\einf$. 
This is not the case for variable opening-angle jet models because of
the extra degree of freedom provided by the distribution of opening
angles.  Thus we find that the Fisher universal jet model considered
here and the variable opening-angle jet model discussed in
\cite{lamb2005b} make different predictions for the distribution in
$\einf$.  Further observations of XRFs can determine this distribution
and therefore distinguish between these two models of jet structure. 
For completeness, we also simulate a variable opening-angle jet model
whose emissivity profile is also a Fisher distribution, and we find
similar results to the universal Fisher jet model.

\section{Simulations}

The Fisher distribution is the only universal jet profile that
satisfies the following two constraints.   Let $\xiso \equiv \ln
\eiso$, and $\muv\equiv\cos\thetav$, where $\thetav$ is the angle
between the line-of-sight and the center of the jet.  Observations tell
us that roughly $dN/d\xiso = a_{1}$, where $a_{1}$ is some constant. 
By the definition of $\thetav$ we know that $dN/d\muv = a_{2}$, where
$a_{2}$ is another constant.   We can describe a universal jet as an
arbitrary function of $\muv$: $\xiso=f(\muv)$.  We wish to choose
$f(\muv)$ so as to satisfy these constraints.

Note that
\beq
a_{2} = \frac{dN}{d\muv} = \frac{dN}{d\xiso} \cdot \frac{d\xiso}{d\muv},
\eeq
and therefore $df/d\muv = a_{2}/a_{1}$, integrating this expression
gives the Fisher distribution
\beq
\eiso = 4\pi A \cdot e^{B \cdot \cos \thetav}
	= 4\pi \hat{A} \cdot e^{(\cos \thetav -1)/\thetan^{2}},
\eeq
where $B=\thetan^{-2}$.  In the small $\thetav$ limit, this reduces to a
Gaussian jet, $\eiso = 4\pi \hat{A} \cdot e^{\thetav^{2}/2\thetan^{2}}$.

Integrating the emissivity over the jet gives $\egt$,
\beq
\egt = 2 \cdot 2\pi A \int_{0}^{\pi/2} e^{B \cdot \cos\theta} 
	\; \sin\theta d\theta
	= \frac{4\pi A}{B}(e^B-1).
\eeq
In this work we consider two oppositely directed jets, hence the
leading factor of 2.  We note that for any non-uniform jet this
quantity is {\it not} the same as the $E_{\gamma}$ inferred using the
method outlined by \cite{frail2001}.  That quantity we term,  $\einf =
\eiso \cdot (1-\cos \thetaj)$,  where $\thetaj={\rm
max}(\thetan,\thetav)$ \cite{dz2004,kumar2003}.  For $\thetan = 0.1$
rad, the quantities $\eiso$ and $\einf$ vary over domains of $\sim 43$
and $\sim 40$ decades, respectively, although observational selection
effects will truncate both of these distributions.  We perform Monte
Carlo simulations using the method presented in \cite{lamb2005b}, using
the detector thresholds from the WXM on HETE-2.

\section{Results}

We consider 3 models.  (1) A Universal Fisher Jet model with $\log\egt
= 51.1$ and $\thetan$ values drawn from a log-normal distribution with
width $0.2$ and $\log \thetan^{0} = -1.0$ (UFJ1), following the
parameters of \cite{dz2004,zhang2004}, (2) a Universal Fisher Jet model
with $\log \thetan^{0} = -1.3$ and $\log\egt = 51.8$ (UFJ2), and (3) a
Variable Opening-Angle Fisher (VOAFJ) jet model with $\log\egt = 51.5$
and $\thetan$ values drawn from a power-law with index $\alpha_{PL} =
-3.3$ and extending for two decades from a maximum of $\pi/2$.  

Figures 1 and 2 show the distribution of detected and non-detected
bursts in various planes for our 3 models.  All 3 models exhibit
roughly equal numbers per logarithmic decade in $\eiso$. For the
universal models, this is a natural consequence of the Fisher profile
of the jet.  For the VOAFJ model this is a consequence of our choice of
$\alpha_{PL} = -3.3$.  We note that the UFJ1 model is unable to
accomodate the highest observed values of $\eiso$ and $\ep$.  The
maximum $\eiso$ generated by a Fisher jet is approximately $\eiso^{\rm
max} = \egt/\thetan^{2}$, which gives $\sim 1.2 \times 10^{53} {\rm
erg}$ for UFJ1 and $\sim 2.5 \times 10^{54} {\rm erg}$ for UFJ2.  

\begin{figure}[htb]
\begin{center}
\rotatebox{270}{\resizebox{3.5cm}{!}{\includegraphics{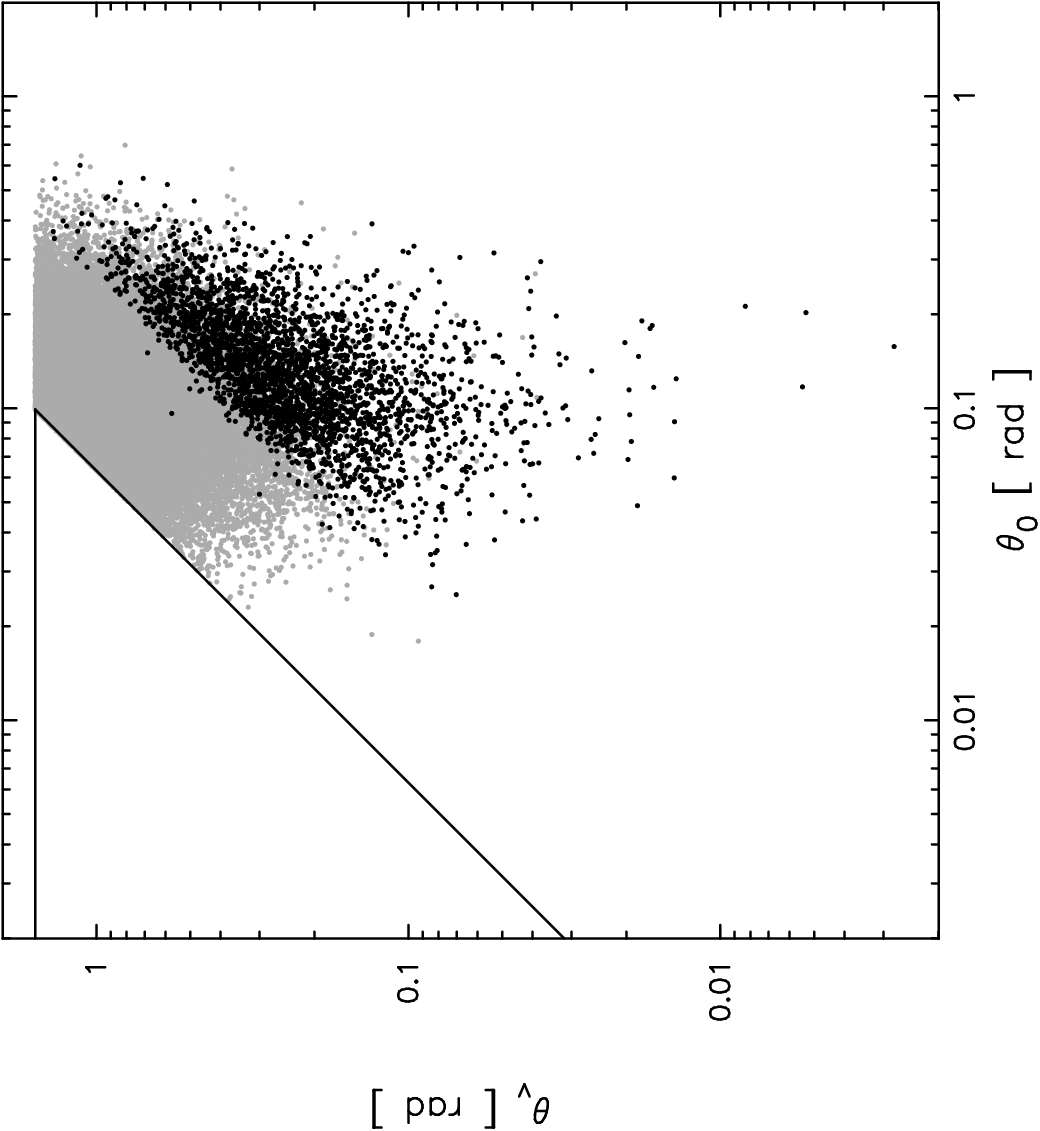}}}
\rotatebox{270}{\resizebox{3.5cm}{!}{\includegraphics{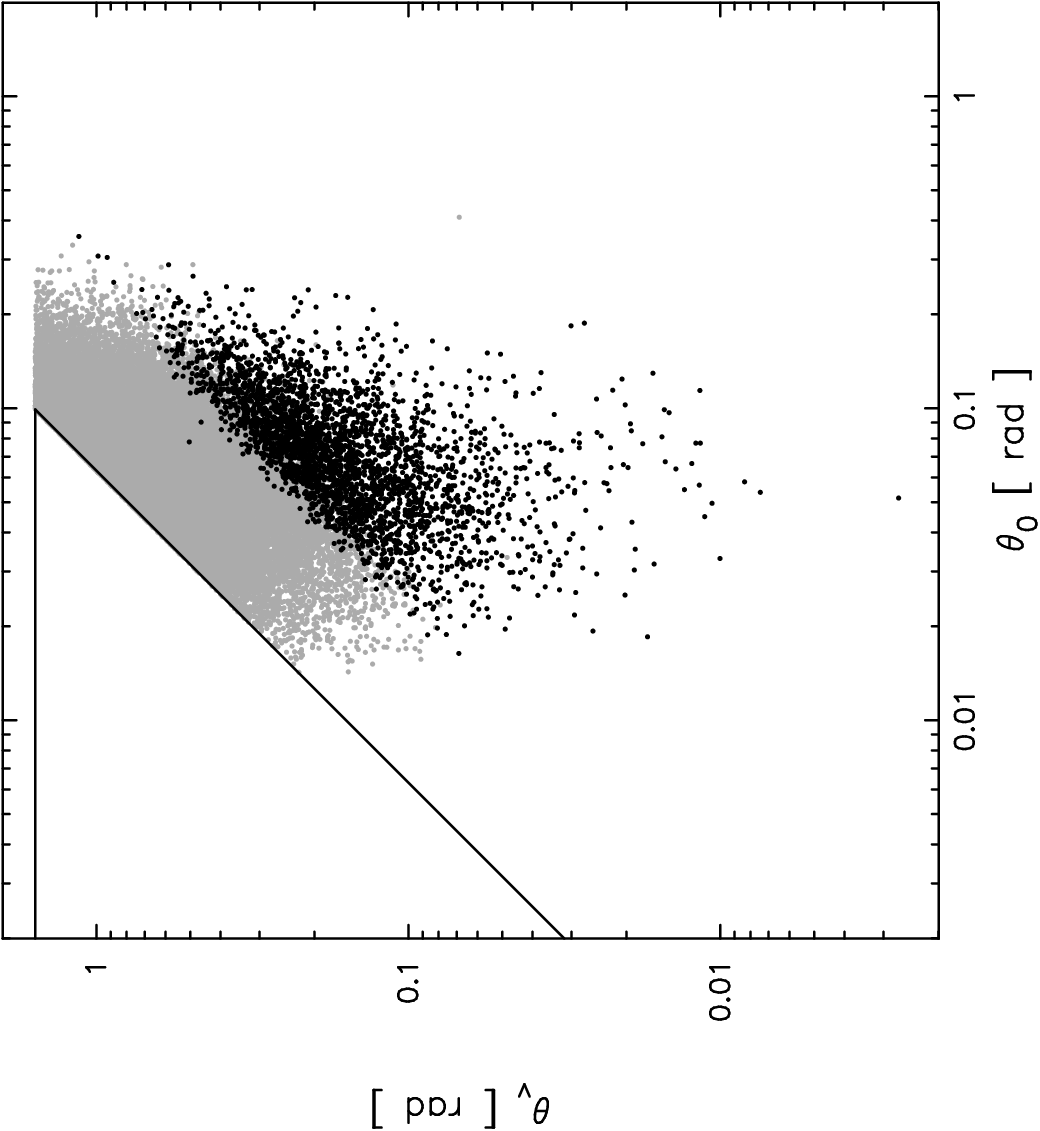}}}
\rotatebox{270}{\resizebox{3.5cm}{!}{\includegraphics{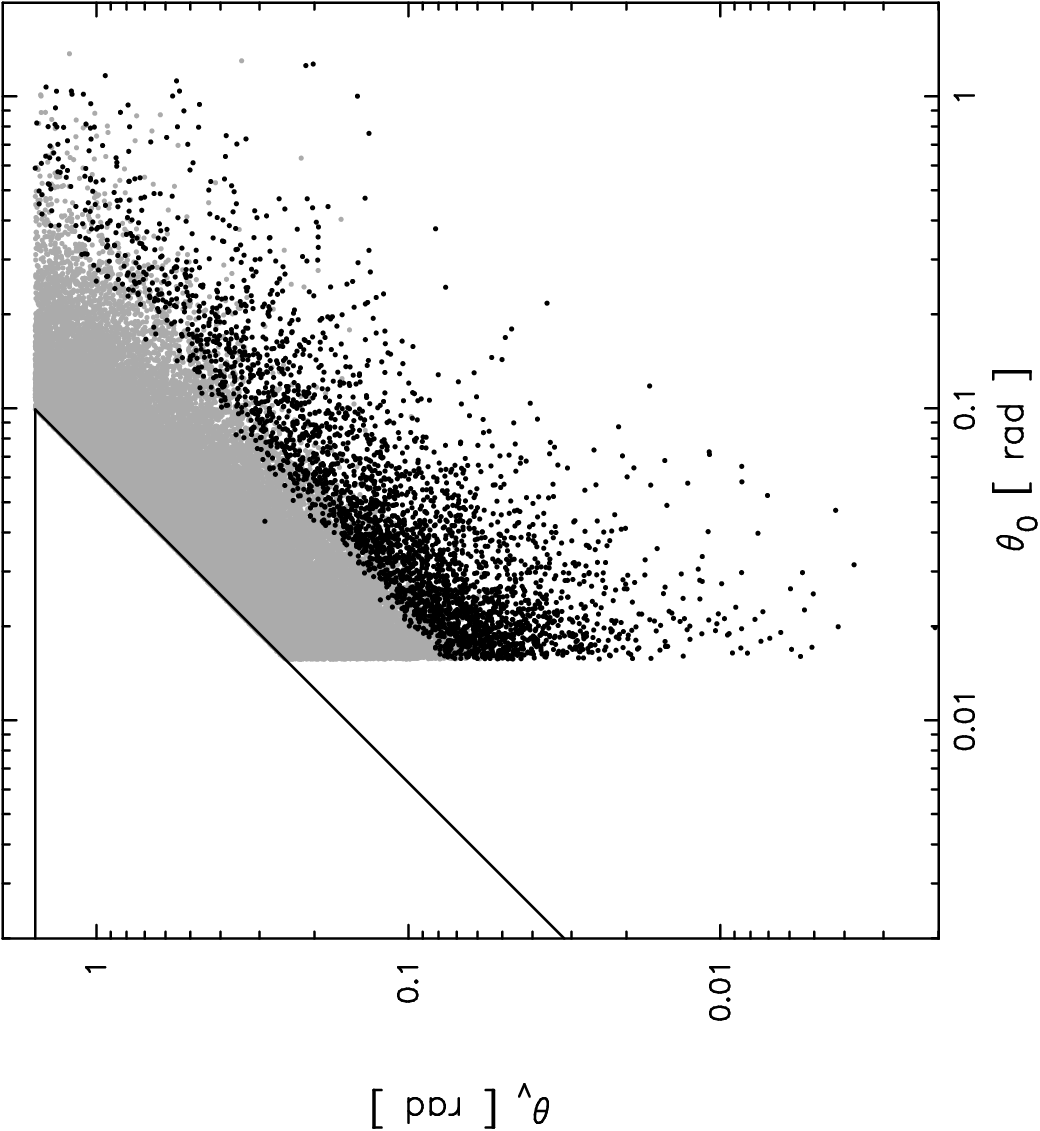}}}
\end{center}
\begin{center}
\rotatebox{270}{\resizebox{3.5cm}{!}{\includegraphics{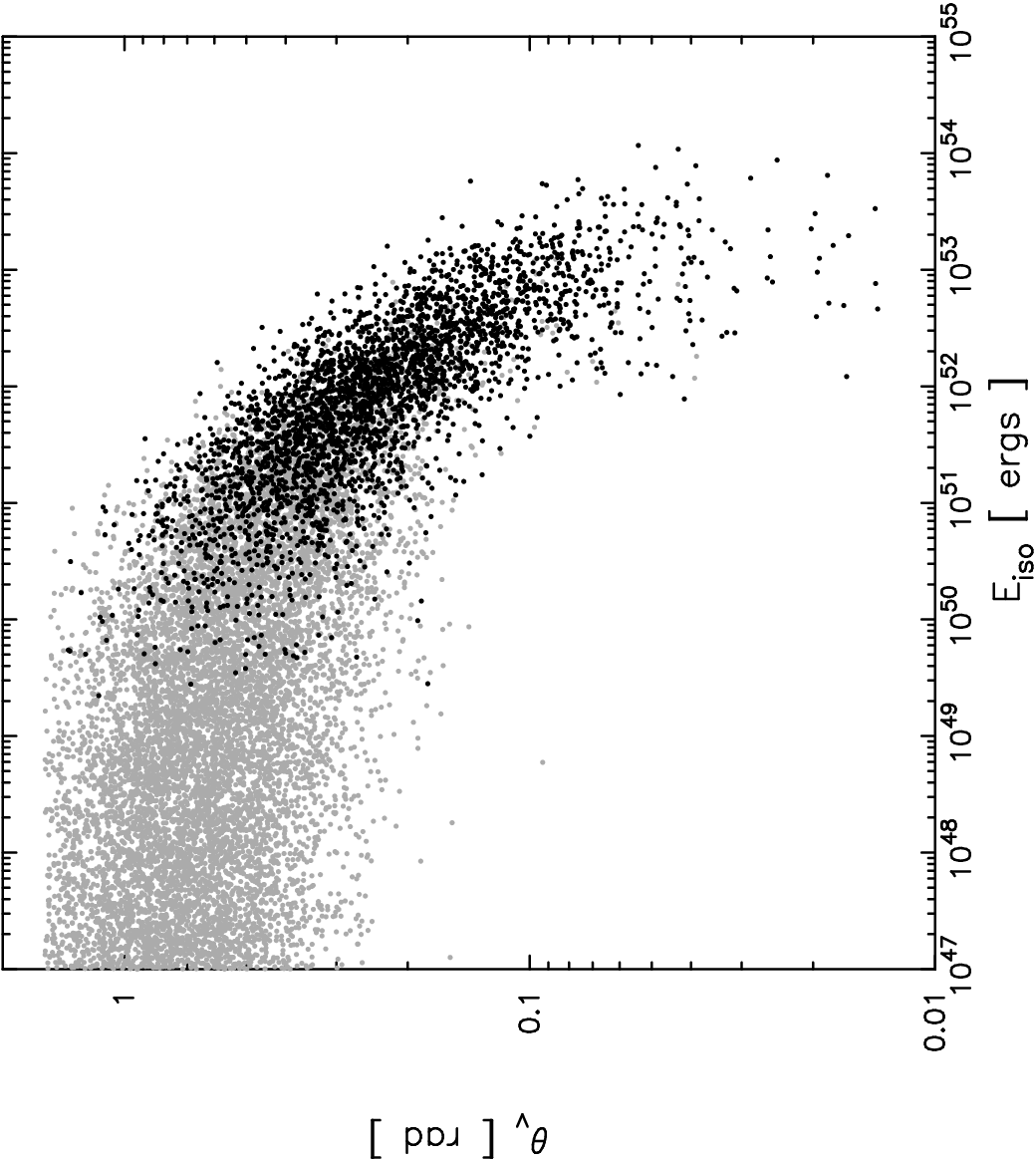}}}
\rotatebox{270}{\resizebox{3.5cm}{!}{\includegraphics{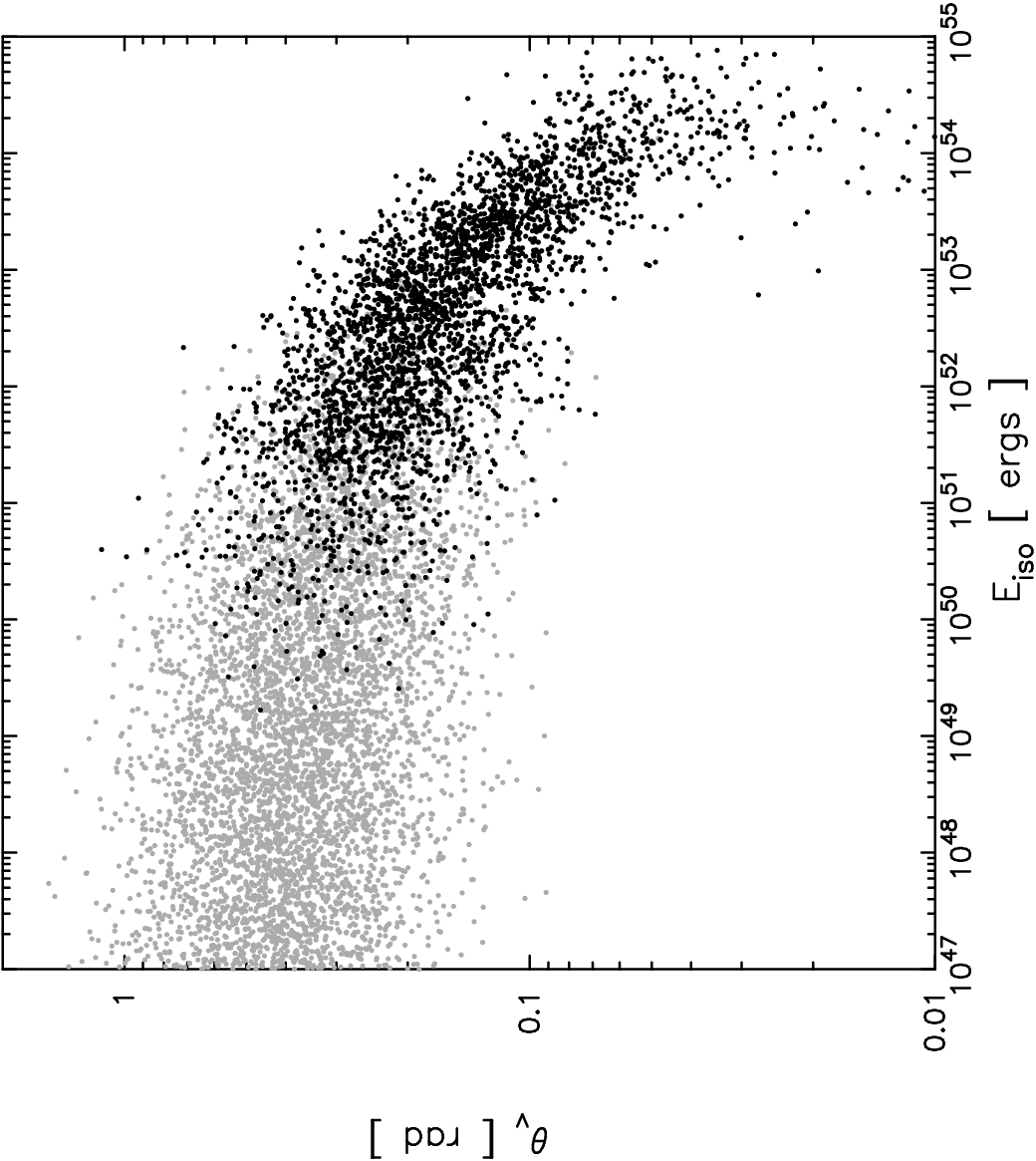}}}
\rotatebox{270}{\resizebox{3.5cm}{!}{\includegraphics{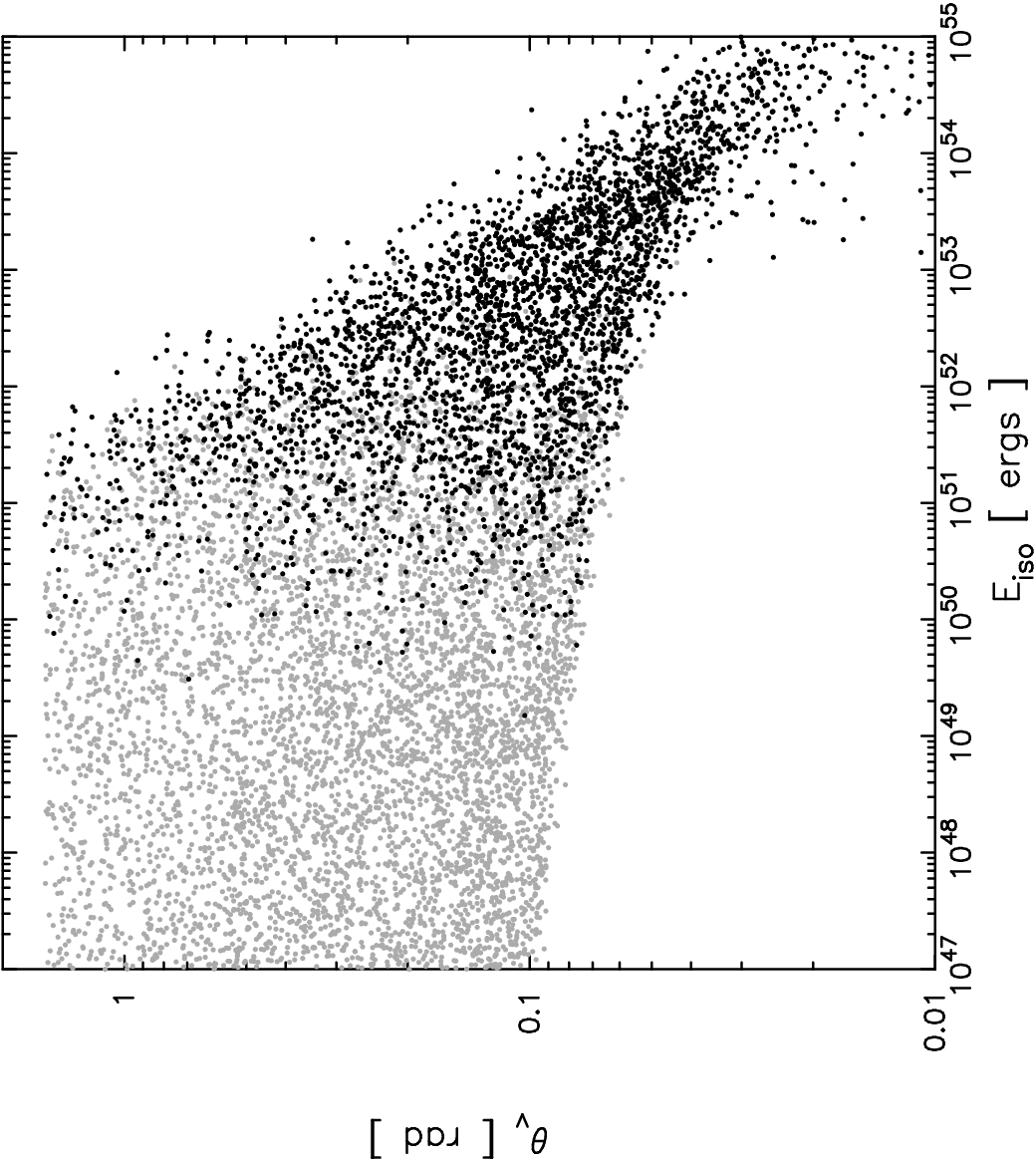}}}
\end{center}
\caption{Scatter plots of detected (black) and undetected bursts (gray)
in the [$\thetan$,$\thetav$] plane (top row) and [$\eiso$,$\thetav$]
plane (bottom row) for our 3 models: UFJ1 (left), UFJ2 (middle) and
VOAFJ (right).  The triangular region in the upper-left corner
represents bursts that we do not simulate to increase the percentage of
detected bursts in a sample of 50,000.}
\end{figure}

The bottom row of Figure 2 shows the histogram of $\einf$ values for
the detected bursts.  It is clear that the $\einf$ distribution does
not agree with the inputted $\egt$ distribution.  For example, in the
UFJ1 model, the peak of the $\egt$ distribution was chosen to
correspond to the ``standard energy'' found by
\cite{frail2001,bloom2003}, however the model is unable to recover that
value using their method.  The $\einf$ distribution typically peaks at
a lower energy and has a tail extending to even lower energies.  It may
be the case that the observed distribution of $\einf$ values does
extend down to lower energies.

\begin{figure}[htb]
\begin{center}
\rotatebox{270}{\resizebox{3.5cm}{!}{\includegraphics{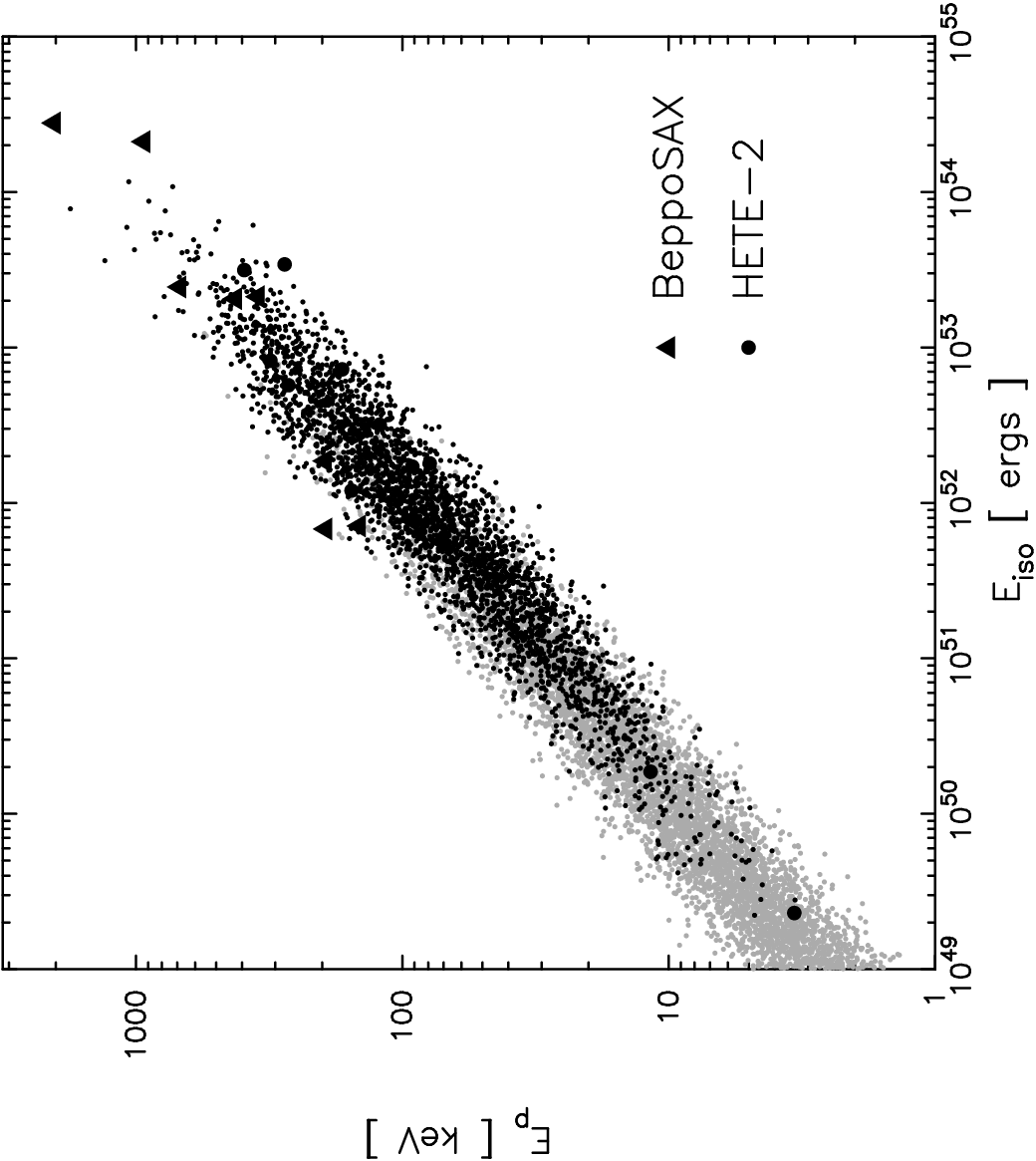}}}
\rotatebox{270}{\resizebox{3.5cm}{!}{\includegraphics{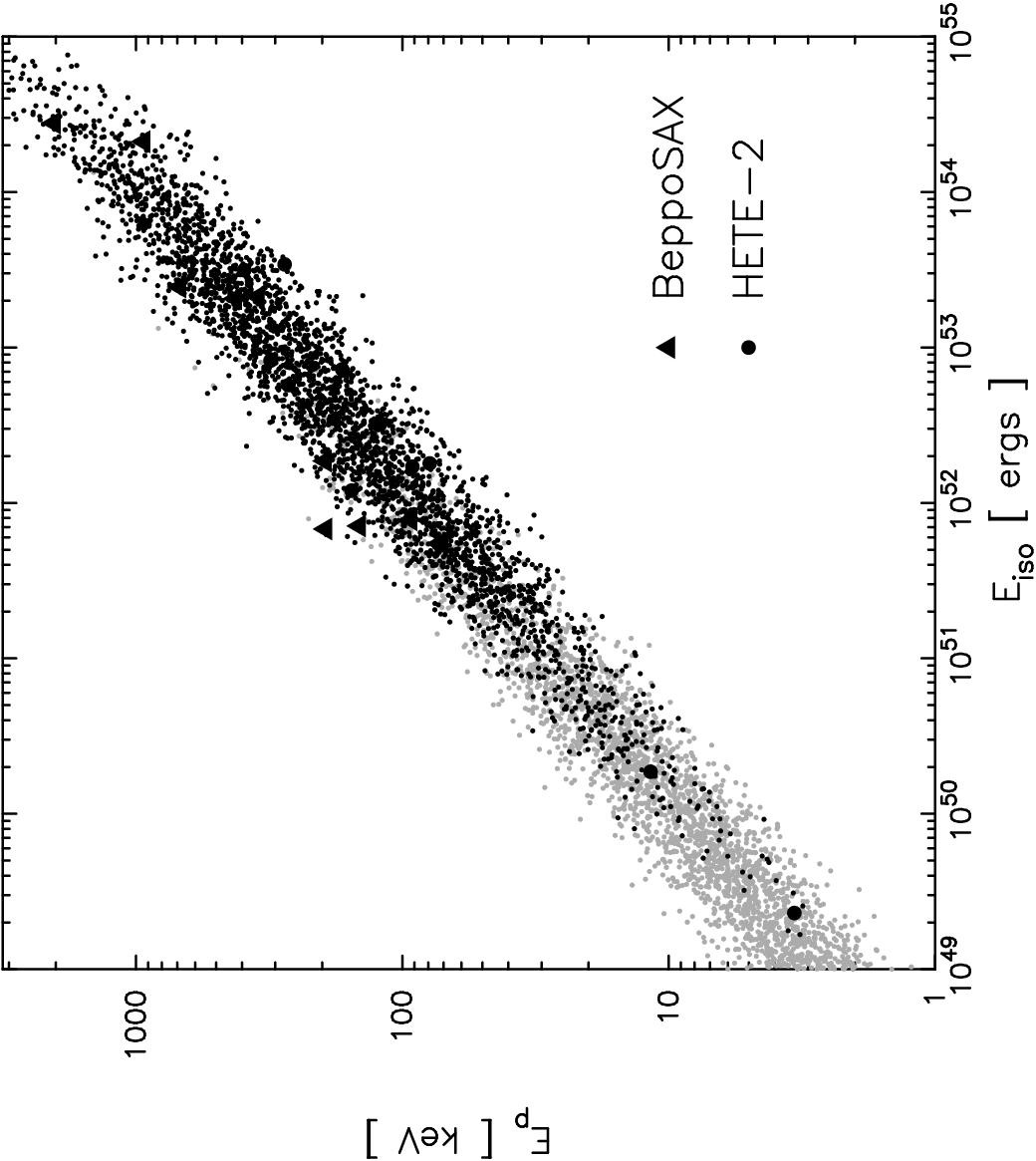}}}
\rotatebox{270}{\resizebox{3.5cm}{!}{\includegraphics{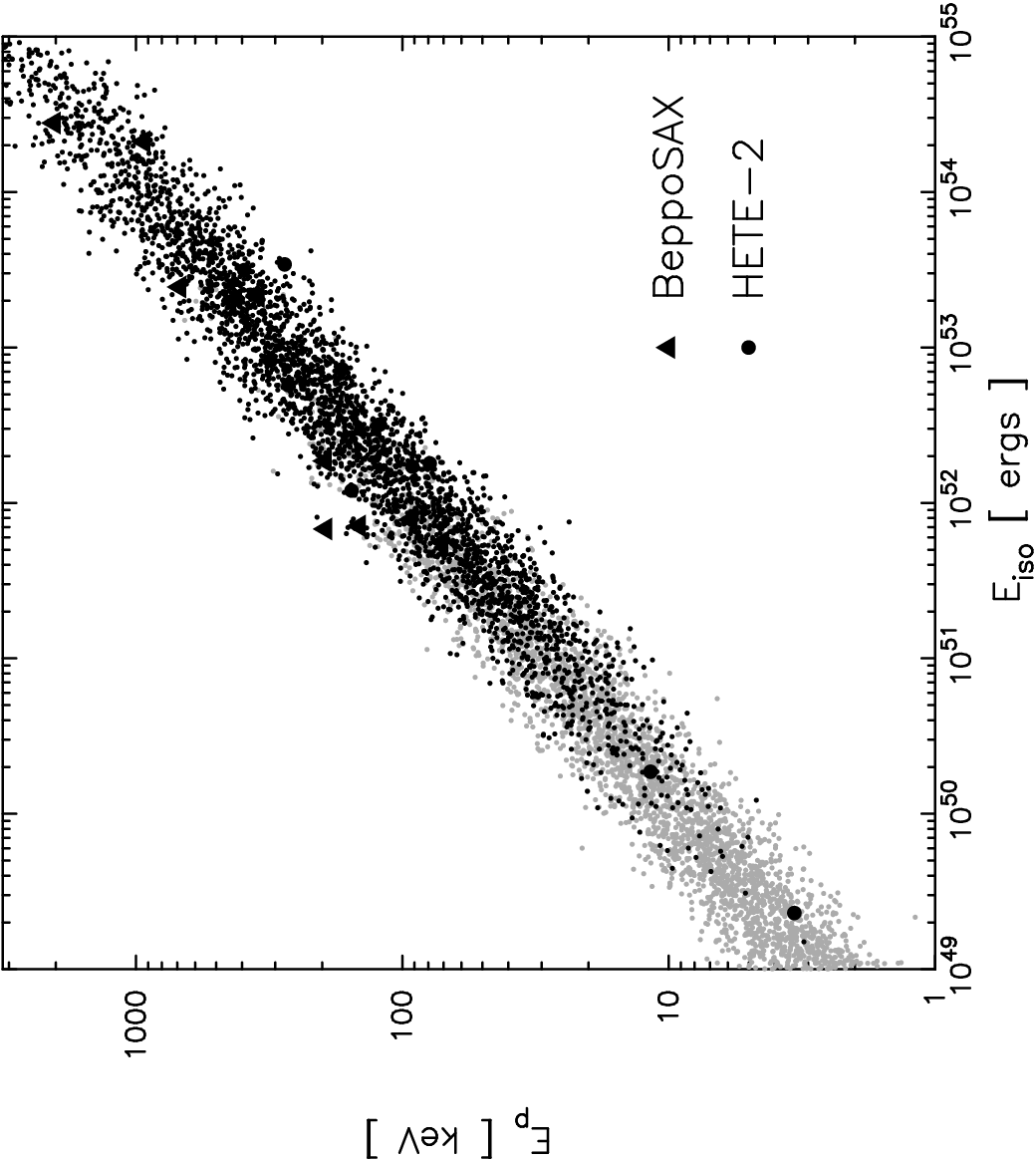}}}
\end{center}
\begin{center}
\rotatebox{270}{\resizebox{3.5cm}{!}{\includegraphics{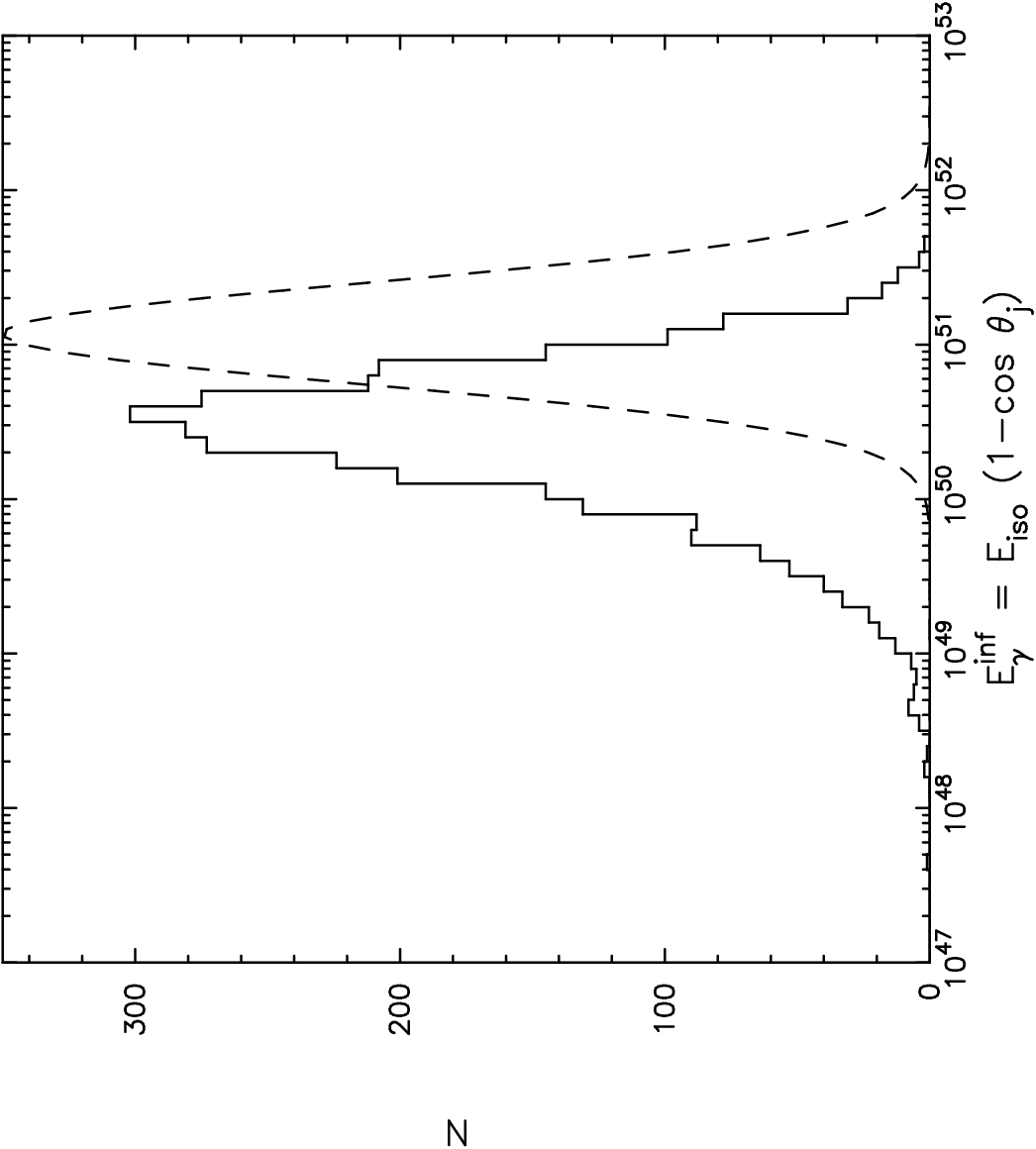}}}
\rotatebox{270}{\resizebox{3.5cm}{!}{\includegraphics{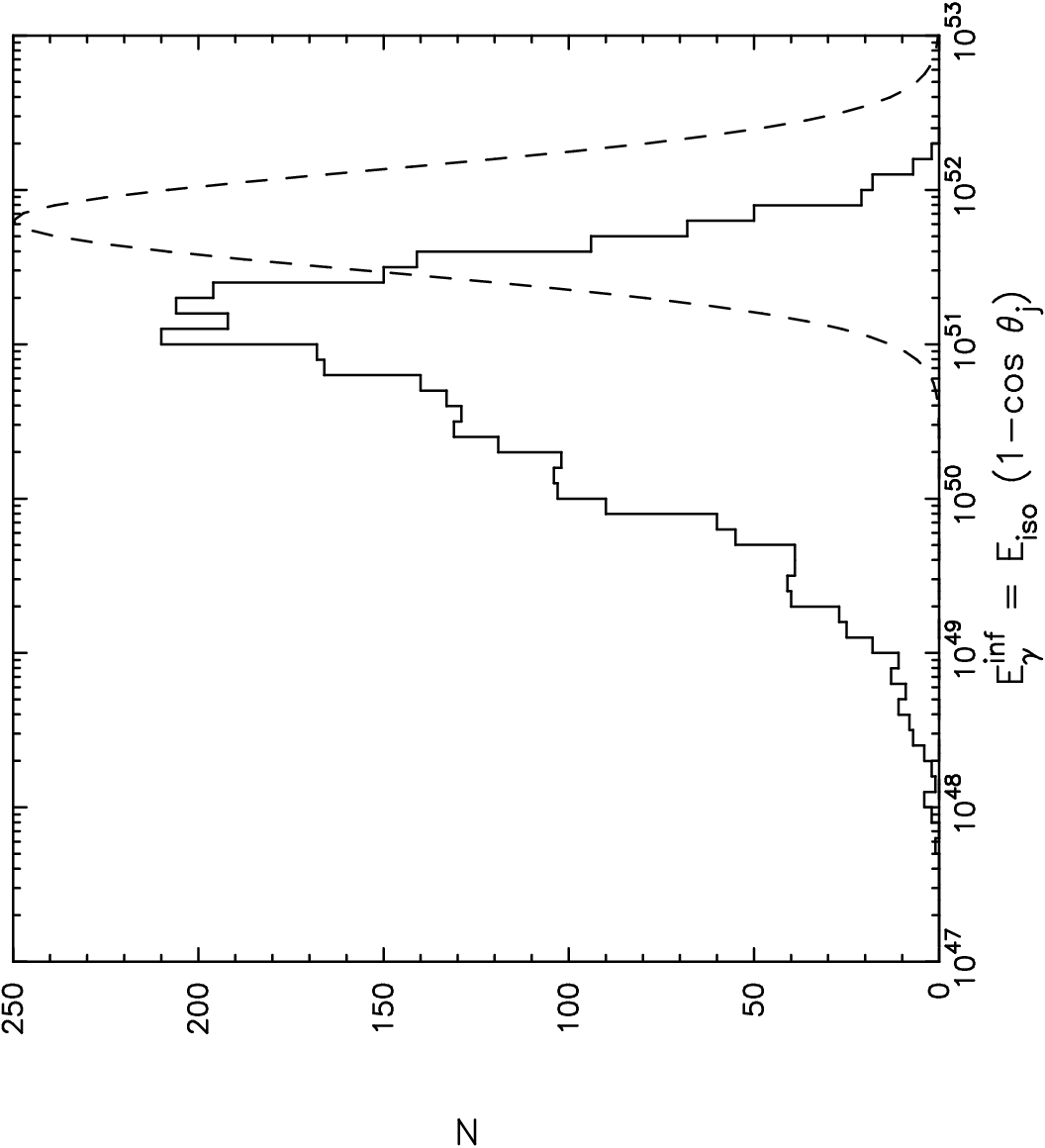}}}
\rotatebox{270}{\resizebox{3.5cm}{!}{\includegraphics{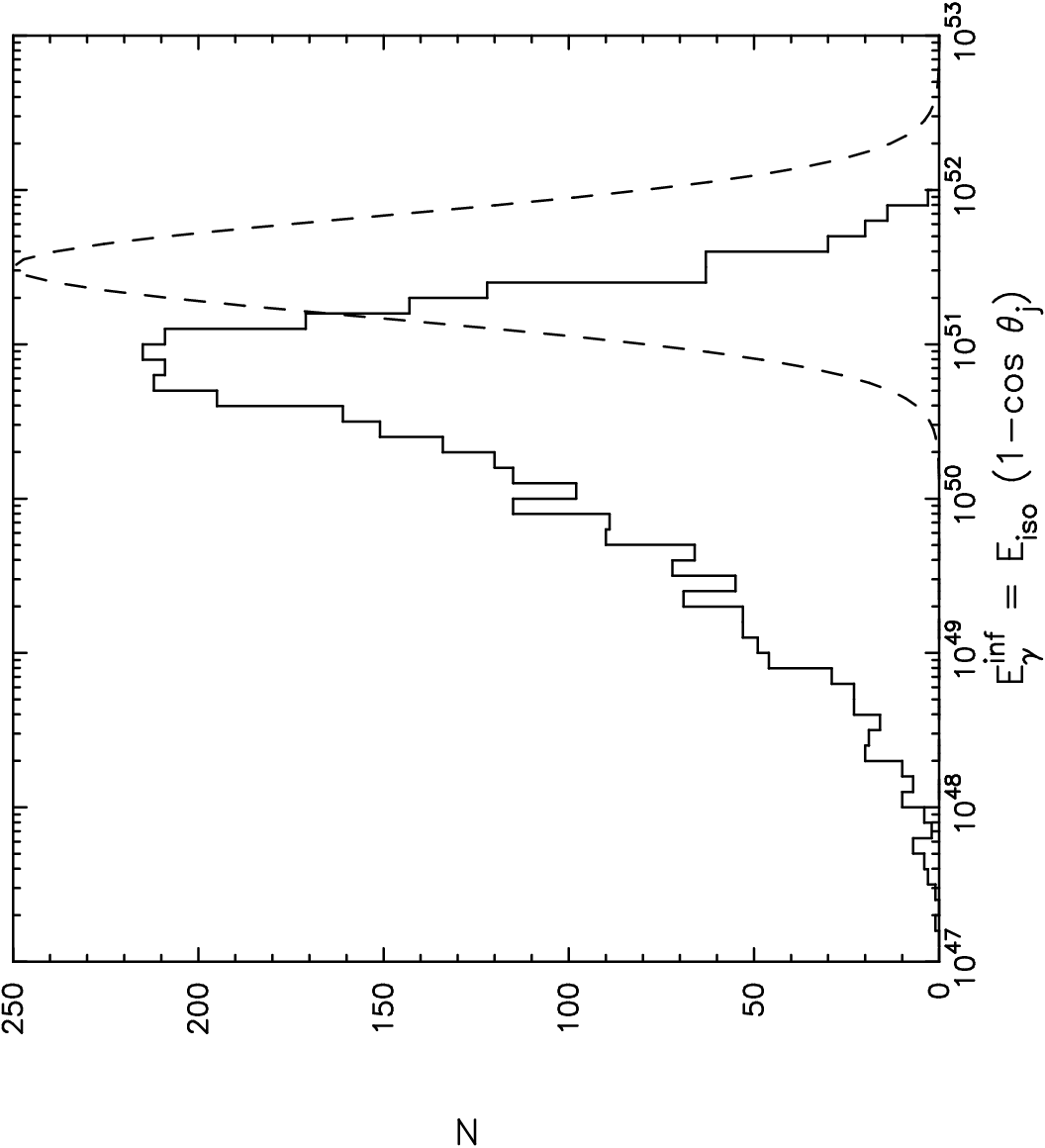}}}
\end{center}
\caption{Top row shows scatter plots of detected (black) and
undetected bursts (gray) in the [$\eiso$,$\ep$] plane for our 3 models:
UFJ1 (left), UFJ2 (middle) and VOAFJ (right).  Bottom row shows
histogram of $\einf$ values for detected bursts, as compared to the
input distribution of $\egt$ (dashed curve).}
\end{figure}

\section{Conclusions}

Both universal and variable Fisher jet models can be found that
reproduce most of the observed properties of XRFs and GRBs.  To
accomodate the highest $\eiso$ bursts, very small jet opening-angles
($\sim 2^{\circ}-3^{\circ}$) may be required in both the variable
opening-angle uniform jet models and in the Fisher models.  Finally,
$\einf$ may be a powerful probe of jet structure, as various models
give different predictions for its distribution.  More observations of
XRFs with redshifts and jet-break times are crucial to answering this
question, highlighting the importance of continuing HETE-2 during the
{\it Swift} mission.


\vspace{-0.2truein}
\begin{thebibliography}{0}

\bibitem{amati2002} \BY{Amati, L., et al.}
  \IN{A\&A}{390}{2002}{81}.

\bibitem{bloom2003} \BY{Bloom, J., Frail, D. A., \& Kulkarni, S. R.}
  \IN{ApJ}{588}{2003}{945}.

\bibitem{dz2004} \BY{Dai, X. \& Zhang, B.} ApJ, submitted, 2004 (astro-ph/0407272).

\bibitem{frail2001} \BY{Frail, D., et al.}
  \IN{ApJ}{562}{2001}{L55}.

\bibitem{heise} \BY{Heise, J., et al.} 2000, in Proc. 
	\TITLE{2nd Rome Workshop: Gamma-Ray Bursts in the Afterglow Era}, 
	eds. \NAME{E. Costa, F. Frontera, J. Hjorth} (Berlin: Springer-Verlag), 16.

\bibitem{kippen2002} \BY{Kippen, R. M., et al.} 2002, in Proc. 
	\TITLE{Gamma-Ray Burst and Afterglow Astronomy}, 
	AIP Conf. Proceedings 662, eds. \NAME{G. R. Ricker \& R. K. Vanderspek}
	(New York: AIP), 244.

\bibitem{kumar2003} \BY{Kumar, P. \& Granot, J.}
  \IN{ApJ}{591}{2003}{1075}.

\bibitem{lamb2005} \BY{Lamb, D. Q., et al.} ApJ, submitted, 2005.

\bibitem{lamb2005b} \BY{Lamb, D. Q., Donaghy, T. Q., and Graziani, C.}
  \IN{ApJ}{620}{2005}{355}.

\bibitem{sakamoto2004b} \BY{Sakamoto, T., et al.} ApJ, submitted, 2004b (astro-ph/0409128).

\bibitem{zhang2002} \BY{Zhang, B., \& Meszaros, P.}
  \IN{ApJ}{571}{2002}{876}.

\bibitem{zhang2004} \BY{Zhang, B., Dai, X., Lloyd-Ronning, N. M., \& Meszaros, P.}
  \IN{ApJ}{601}{2004}{L119}.

\end{thebibliography}
\end{document}